# Drop Impact Printing

Chandantaru Dey Modak[1], Arvind Kumar[1,2], Abinash Tripathy[1,3] and Prosenjit Sen[1,*]

[1]Centre for Nano Science and Engineering, Indian Institute of Science, Bangalore, India, 560012

**Corresponding Author's Email:** prosenjits@iisc.ac.in

[2]Present Affiliation - Australian Institute of Bioengineering and Nanotechnology, The University of Queensland, Brisbane, Australia, 4072.

[3]Present Affiliation - Department of Mechanical and Process Engineering, ETH Zurich, Sonneggstrasse 3, CH-8092 Zürich, Switzerland.

## TOC

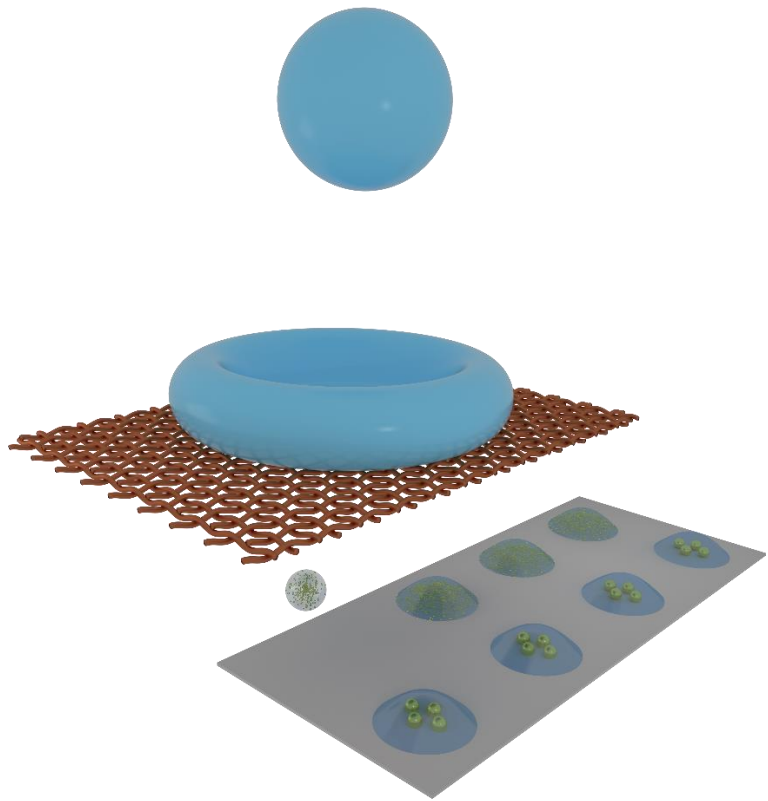

## Abstract

Hydrodynamic collapse of a central air-cavity during the recoil phase of droplet impact on a superhydrophobic sieve leads to satellite-free generation of a single droplet through the sieve. Two modes of cavity formation and droplet ejection was observed and explained. The volume of the generated droplet scales with the pore size. Based on this phenomenon, we propose a new drop-on-demand printing technique. Despite significant advancements in inkjet technology, enhancement in mass-loading and particle-size have been limited due to clogging of the printhead nozzle. By replacing the nozzle with a sieve, we demonstrate printing of nanoparticle suspension with 71% mass-loading. Comparatively large particles of 20µm diameter were dispensed in droplets of ~80µm diameter. Printing was performed for surface tension as low as 32 mN/m and viscosity as high as 33 mPa·s. In comparison to existing techniques, this new way of printing is widely accessible as it is significantly simple and economical.

**Keyword**: Droplet impact, Superhydrophobic, Sieve, Satellite-free, Drop-on-demand.

# Introduction

Dispensing small droplets is of great research interest because of its numerous applications in electronic industry, medical science, automobiles, rapid prototyping etc. However, printing small droplets is challenging due to the dominance of surface tension at length scales smaller than the capillary length $\sqrt{\gamma/\rho g}$ (where $\gamma$ is surface tension, $\rho$ is density and $g$ is gravitational acceleration). For generating small droplets, surface tension force is usually overcome by applying an external force (e.g., electrical, thermal or acoustic). Inkjet printers are well established for conventional printing. However, newer applications require printing of more exotic inks containing biological samples, biopolymers, micro/nano particles, etc. Since conventional inkjet technology is not designed to work with newer inks, it fails to provide the desired resolution, accuracy and widespread applicability[1–5]. This has led to the emergence of other printing techniques[6–9] using acoustic[10], electrohydrodynamic[11–13], laser-assisted[14] or microfluidics[15] based designs. Use of complex technology in these techniques prohibitively increases their setup and operational cost[16,17]. Hence, the availability of these printing techniques has been restricted to a selected community only.

Most micro-droplet printing technologies use a nozzle based dispensing configuration[17] with complicated actuators[18–20]. The nozzle primarily focuses the applied force and hence determines the ejected droplet size. In these technologies, two main disadvantages of satellite droplets[21–23] and nozzle clogging[9,24,25] mostly remains unaddressed. Satellite drops are unwanted products of the droplet formation process which reduces pattern quality. Nozzle clogging predominantly happens due to solvent evaporation while attempting to print inks with either higher mass-loading or large particles. Nozzle clogging is mostly destructive requiring replacement of the expensive nozzle. These issues severely restrict nozzle based printing to use liquids with either limited range of properties[22,23] or definite mass loading[9,24–26].

This report describes a new way of printing that mitigates these issues. Based on impact of a droplet on a superhydrophobic sieve, the setup is exceptionally simple. Cavity collapse during the recoil phase leads to satellite free generation of a single micro-droplet. The printing technique outperforms conventional inkjet technique in most aspects. By replacing the nozzle with a sieve, we demonstrate printing with high mass-loading (71%) and large particle size (20µm). Apart from this, the technique is cost-effective, compact in size, easy to operate and allows instant reconfiguration for different micro-droplet sizes. Using this setup, the manuscript reports printing of various inks for different applications. In addition to traditional applications this technique can be used for: (1) ceramics-based 3D printing[27] for dental prostheses[28], architectural modeling[29], etc.; (2) dispensing biological samples for single cell applications, 3D organ printing, etc.[30], and; (3) printing for electronic applications[5,9]. Apart from its versatility, this technique is remarkably affordable and hence will make drop-on-demand printing widely accessible.

## Results and Discussion

### Cavity Collapse Driven Single Micro-Droplet Ejection

Outcome of drop impact on a superhydrophobic sieve is determined by the balance between the dynamic pressure ($\sim \rho U_0^2$) of the impinging droplet and the breakthrough pressure ($\sim 4\gamma/L$) of the sieve [31,32]. Here, $U_0$ is the impact velocity, $\rho$ is the density, $\gamma$ is surface tension and $L$ is the size of the pore as shown in Figure 1(a). In our impact experiments, water droplets of diameter, $D_o$ ~2.56 mm were released from different heights varying from 2 cm to 5 cm. Droplet ejection was captured using a high-speed camera (Photron FastCam) operating at frame rates as high as 75,000 frames per second. Using superhydrophobic sieve #0.009 (refer to Supplementary Table S1 for geometrical parameters), possibility of single droplet generation was evaluated. At lower impact velocities ($U_0 = 69$ cm/s, $We = 17$) the liquid failed to penetrate the mesh (Supplementary Figure S1(a)) as the breakthrough pressure was higher than the dynamic pressure.

As the impact velocity increased, a regime of single droplet ejection was observed (Video SV1 and Supplementary Figure S2). However, the micro-droplet creation was not observed during the impact phase. The impact pressure ($\sim \rho U^2$) was not enough for ejection of liquid through the pore and its subsequent separation by Rayleigh-Plateau instability[33,19]. Only during the recoil phase micro-droplet generation was observed. Hence, this phenomenon has been termed as recoil ejection[31]. Without identifying a physical cause, prior literature has attributed this ejection to increase in local pressures during the retraction phase[31]. On further increasing the impact velocity ($U_0 = 83$ cm/s, $We = 25$), micro-droplet ejection was observed during the spreading phase. This however led to generation of multiple droplets (Supplementary Figure S1(b)). Henceforth, we focus on the single droplet generation by recoil ejection.

Experimentally, we observed formation of an air-cavity during interface retraction and its collapse just prior to the recoil ejection as seen in Figure 1(b). Formation of air-cavity has been previously reported for impact on flat hydrophobic surface[34]. Droplet impact creates capillary waves, which leads to formation of a cylindrical air-cavity trapped between the retracting interface. Motion of the interface causes the cavity to collapse and the kinetic energy of the fluid converges along the axis of collapse. This inertial focusing causes the interface velocity to diverge[35]. Local dynamic pressure ($\sim \rho U^2$) at the collapsing front becomes much larger than the impact dynamic pressure ($\sim \rho U_0^2$). On flat surfaces, the resulting hydrodynamic singularity causes ejection of a narrow high-speed jet[34]. For impact on superhydrophobic sieves, the pore limits the lateral extent of the collapsing cavity. It also sets the lateral boundary for the interface motion resulting from the cavity collapse. Sieves topography changes the collapse dynamics and a single micro-droplet is ejected as seen in Video SV2 & Supplementary Figure S3(a),(b). This mode where the cavity is formed during the initial impact has been termed impact cavity (IC) mode.

In our experiments with sieve #0.009, we observe a new mode of cavity formation as shown in Video SV3 & Figure 1(c). The liquid penetrating the meshes during impact is observed to recoil back and move up[36]. This is due to the surface energy stored in the penetrating liquid ($\sim \gamma L^2$). The liquid moving up from the pores completely fills the initial cavity (impact cavity) formed during the spreading phase. Interestingly, the interface recoiling from the pores does not stop at the top surface of the mesh. The interface is observed to move up through the droplet and a new cavity is formed. This cavity has been termed as recoil cavity (RC). Collapse of cavities formed by both IC and RC mode leads to single droplet generation which we use for printing application.

**Satellite Free Droplet Ejection**

Satellite drops are an artifact of breaking an ejected stream into droplets due to Rayleigh instability. Hence, a common strategy for eliminating satellite droplets has been to attain separation with shorter neck lengths. Nozzle-based printers commonly use actuation waveforms with positive and negative pressure pulses. An initial positive pulse is used to eject the liquid, whereas the negative pulse pulls back the bulk-liquid to enable quick separation[37]. Conceptually these schemes attempt to create a short pulse of focused energy at the tip.

In recoil ejection we naturally observe satellite free droplet creation. Here, the collapse of the cavity focuses the kinetic energy. The pore limits the resultant interface motion and distributes this energy over a length scale of the pore ($\sim L$). Beyond this length scale, the dynamic pressure quickly falls to bulk values (Supplementary Figure S4). The droplet separation is further aided by the bulk flow which during the collapse (recoil phase) is pointed away from the surface as observed in simulation results (see Supporting Information). In conjunction both these effects inherently create the conditions generated by the complex pulse train in nozzle-based printers. The importance of recoil ejection for satellite free droplet generation became apparent when hydrophobic meshes were used. In hydrophobic meshes droplets were generated in impact ejection mode only. In this mode, where inertial focusing is absent, longer necks and satellite droplets were observed.

**Drop-on-Demand Printing**

Water droplets of different diameter were dispensed using sieves with different pore openings (see Figure 1(e) for SEM images of the sieve). Supplementary Table S2 shows the range of dimensionless number for which water droplet printing was carried out with varying pore openings. Figure 1(d) shows the plot of ejected droplet diameter (measured using ImageJ[38]) as a function of pore opening. Size of the ejected droplet was proportional to the pore opening, $D_p = 0.88 * L^{1.07}$ (except for sieve #0.012). For sieves other than #0.012, the liquid from initial penetration was able to retract back and the whole microdroplet volume was from the liquid penetrating the mesh after the cavity collapse. We name it Collapse Penetration Mode (CPM). Different possible outcome of droplet impact is shown in Supplementary Figure S5.

Compared to other sieves, sieve #0.012 ejects out higher droplet volume for its pore opening. Sieve #0.012 has the largest pore opening. Unlike other meshes, the liquid from initial (impact) penetration is unable to retract back (Supplementary Figure 6(a)-(d)). This liquid combines with the liquid brought in by the cavity collapse during recoil and leads to a higher ejection volume (Supplementary Figure 6(e),(f), Video SV4). This mode of micro-droplet creation has been named Impact Penetration Mode (IPM). Although the droplet volume for sieve type #0.012 is bit higher as compared to other sieves, it still gives satellite free dispensing. Thus, this technique provides the capability of printing a wide range of single droplet of diameter ranging from 94 µm to 926 µm (Video SV5).

Capability to print a broad range of liquids was validated by using Newtonian (other than water), and non-Newtonian fluids of varying viscosities and surface tensions. Viscosity was varied by adding glycerol to water. Surface tension variation was obtained by adding PEG or ethanol to water. Supplementary Table S3 shows the properties of these liquids. As seen in Figure 2(a),(b), single droplet printing was possible for viscosity as high as 33 mPa·s and surface tension as low as 32 mN/m. The ejected droplet diameter was mostly independent of varying viscosity (Figure 2(a)) and surface tension (Figure 2(b)). However, transition from CPM mode to IPM mode of ejection led to a slight increase in volume. For the largest pore opening (sieve #0.012), IPM mode of droplet creation was observed for all values of surface tension and viscosity. For other meshes, transition to IPM mode of ejection was observed at higher viscosities. Similarly, a transition from CPM mode to IPM mode of ejection was observed for lower surface tension values. Finally, droplet impact was used to print viscoelastic liquid (Xanthum gum and water) of varying concentrations (1-10%, v/v). Single droplet printing was observed up to viscosity 20mPa.s (Supplementary Figure S6).

Figure 2(c) shows the printable region for Newtonian fluid in the terms of Ohnesorge number ($Oh = \mu/\sqrt{\rho\gamma L}$, $\mu - Viscosity$) and Reynolds number ($\frac{\rho U_0 D_0}{\mu}$). As compared to traditional drop-on-demand printers, the current technique can print using a wider range of fluid properties [39,40]. Figure 2(d) compares the Z number ($1/Oh$) for our technique with traditional drop-on-demand printers. Drop-impact printer can print for Z values varying from 3 to 200, which is significantly better than the reported range of 1 to 10 for commonly used techniques. Below Z < 3, viscous force is high, so the liquid is unable to penetrate the sieve and maximum Z ~ 200 corresponds to water drop using sieve #0.012.

Which mode (IPM or CPM) is observed is determined by a competition between the different timescales pertaining to droplet impact and liquid penetration. The penetrated interface is able to recoil back if its dynamics is faster (timescale is shorter) than that of the impacting droplet. Impact dynamics of the parent droplet is dominated by inertia with a timescale of $\tau_d \sim \sqrt{\rho D^3/\gamma}$. Timescale of liquid penetration & retraction is determined by liquid inertia and viscosity. In purely inertial regime the timescale of the penetrated interface is given by $\tau_i \sim \sqrt{\rho L^3/\gamma}$. Collapse penetration mode will be observable when the non-dimensional ratio of these timescales (timescale factor $TSF \sim \sqrt{L^3/D^3}$) is smaller than a critical value. In viscous regime, the timescale of the penetrated interface is given by $\tau_v \sim \mu W/\gamma$ (where $\mu$ is viscosity and $W$ is the width of mesh wire). The crossover from inertial regime to viscous regime happens when the timescale to setup viscous flows in the pore ($\sim \rho L^2/\mu$) is smaller than the inertial timescale ($\sqrt{\rho L^3/\gamma}$). This implies viscous effects become dominant above a crossover *Ohnesorge* number $Oh_{cr}$. The equations can be rearranged to get a common timescale factor given by

$$TSF = f\left(\frac{Oh \times \left(\frac{W}{L}\right)}{Oh_{cr} \times \left(\frac{W}{L}\right)_{cr}}\right) \times \sqrt{\frac{L^3}{D^3}} \qquad (1)$$

$$f(x) = 1, for\ Oh < Oh_{cr}$$

$$f(x) = x, for\ Oh \geq Oh_{cr}$$

For calculation of $TSF$, it is necessary to identify the critical $Oh$ beyond which viscous forces are no more negligible. We identify $Oh_{cr}$ by looking at the sieve with the largest pore (#0.009) which transitions form CPM to IPM ($Oh_{cr} \approx 0.03$). The enhanced role of viscosity is also evident from Weber Number required for the ejection of a single droplet (Supplementary Figure S7). $TSF$ is plotted in Figure 1(e). For sieve #0.012, large mesh size leads to IPM ejection even in the inertial regime. For our experiments a critical $TSF$ of 0.04 seems to separate the two regimes well.

**Printing of Large Particles**

In conventional nozzle-based inkjet printers, the nozzle diameter limits the particle size that can be printed. It has been reported that for printing of suspensions, the printer nozzle diameter should be 100 times greater than the particle size, otherwise nozzle clogging may occur[41]. Printing of larger particles is required for cell suspensions, functionalized microbeads, 3D microparticle structuring for dental prosthetics, etc. By eliminating the nozzle, the drop impact printing performed considerably better. Even with sieve #0.0020 having the smallest pore opening of 76.2 µm, we could print 20 µm polystyrene beads without clogging (Refer Supplementary Table S4 for nanoparticle sizes). Figure 3(a) illustrates the broad range of particle size that can be printed using drop impact printing. Capability to handle different particle sizes is quantified as a ratio of nozzle diameter to particle diameter. For drop impact printing this ratio goes down to 4 from the traditional known value of 100. The significant advancement can be attributed to the sieve configuration where the sample liquid is only in intermittent contact (~10 ms) with the nozzle (sieve pore). This eliminates the probability of nozzle clogging due to particle agglomeration. We further quantify the probability of single bead trapping. The probability of getting single 20 µm polystyrene bead in a 0.268 nL volume drop is 32% (Figure 3(b)).

Viscosity of dilute suspensions is known to vary linearly with concentration. However, rheology of suspensions with higher concentration of nanoparticles is complicated due to complex particle-particle and particle-fluid interactions. Rheological behavior of such suspensions is expected to vary not only with mass-loading but also the size of the suspended particles. We have studied the effect of changes in suspended particle sizes on micro-droplet ejection. As seen in Figure 3(c), droplet diameter does not vary with particle size in the suspensions for a given mass loading of 9%.

**Printing with High Mass Loading**

Nozzle clogging also depends on the mass loading. Printing inks with higher mass loading is beneficial as it reduces the number of reprints required for achieving higher thickness. As loading increases, the viscosity increases which makes jetting of suspensions difficult. However, the major challenge of printing suspensions is due the enhanced nozzle clogging from preferential drying of the solvent at the nozzle tip[24]. Previous reports state that the clogging can be reduced by using proper dispersion agent. Even with these measures, printing could be achieved for mass loadings only up to 40%[42,43].

We carried out experiments to estimate the maximum mass loading that can be achieved using sieve #0.009. The ink was formulated using different concentrations of $ZrO_2$ nanoparticle dispersed in 10 vol% polyethylene glycol (PEG). Figure 3(d) shows SEM images of ejected droplet for different mass loadings. The illustration shows the technique range to print high mass loading as compared to others technology. We were able to achieve repeatable micro-droplet generation for a maximum mass loading of 71%. In these experiments the mesh was slightly tilted to ensure that the impacting droplet did not settle on the mesh after impact. Using 71% mass-loading we were able to achieve deposition thickness of 16.9 µm in a single print (Figure 3(e)). As expected, with increase in mass loading we observe an increase in the deposition thickness. The ejected droplet diameter was found to be approximately same with variation in mass loading (Figure 3(f)). At the highest mass-loading small amount of residue was left on the sieve by the impacting droplet. In our experiments, this effected the lifetime of the impact location on the sieve to a limited number of impacts. Unlike clogged nozzle, for our case the residues can be easily removed by washing with a secondary liquid.

**Printing Accuracy**

The printing accuracy was evaluated in terms of droplet size consistency and droplet exit angle. The droplet size accuracy was measured by dispensing an array of 50 droplets of aqueous silver nanoparticles (4% v/v) through sieve number #0.009 and #0.0020. The deposited droplets were heated at 90 °C for 4 hours and the size was measured from optical images using Image J software. As seen in Supplementary Figure S8(a) the deposited droplets are monodispersed with size of 559 μm ±11 μm and 83 μm ±2 μm for sieve #0.009 and #0.0020 respectively.

The droplet exit angle was determined for sieves with different pore openings by using the images extracted from the high-speed videos. The ejection angle was measured with respect to a vertical axis representing the ideal ejection path. The angle of deviation range was measured to be as high as 5° (Supplementary Figure S8(b),(c)). This angle of deviation leads to a displacement error of ~90 μm for a substrate placed 1 mm below the sieve. This deviation is attributed to the lack of alignment between the mesh pore and the impacting droplet.

**Printing for Biological Applications**

In biological science, room-temperature printing of microarrays (bacteria, DNA, cells, proteins etc.) for gene expression analysis, single cell printing for basic biological cells studies, biopolymer printings etc are of paramount interest. The present technique was tested for printing smaller volume of bio samples and molecules. We performed single drop printing of RBC cell suspensions. The RBC cells of varying concentrations were printed on a glass slide. Figure 4(a) shows the printed drop of different cell concentrations. The concentration of the cell solutions was varied from $4*10^4$ to $62*10^4$ cells/mm$^3$. In addition to this, we investigate the number of cells per droplet with varying cell concentrations (Fig. 4(b)). The sample data is for 50 drops for both sieve #0.009 and #0.0045. The analysis revealed that as concentrations increased, the number of cells per droplet increases. Also, the cells remain isolated within the droplets. This gives us the benefit of using very small sample volume that will be isolated from each other and also reduces the time required for pipetting and placing samples. Further the study was extended to print single cell (MDAMB 231) in a single drop of volume 0.268 *nL* (Figure 4(c)). The present technology of printing large cells solutions exists, but the nozzle clogging still remain a major challenge[44,45]. Thus, this technique provides us effective solution for clogging free printing of large cells for different applications.

One of the other ways of doing cell culture-based studies is by patterning. The culture substrate is patterned with different wettability. The simplest way to do is to change the surface chemistry, making superhydrophobic and superhydrophilic arrays. The drop impact printing can also be used for making such gradient surfaces. DMEM liquid was used as printing ink in our case and arrays of DMEM drops were printed on Teflon coated substrate (Figure 4(d.1)). Upon drying the DMEM drops became hydrophilic and the rest of the Teflon coated surface remained hydrophobic, thus making a wettability gradient. When a cell solution was allowed to flow over the gradient surface, the solution was trapped within the hydrophilic area (Figure 4(d.2),(d.3)). Hence this drop impact printing technique provides us a new way to make such gradient surfaces without modulating the surface chemistry.

Additionally, to realize the possibility of printing viscous bio-ink for 3D printing applications, polyacrylic acid was used as a model printing liquid. Polyacrylic acid 1.25% (w/w) was used for printing droplet volume of 0.4 µL (948 µm diameter) in the form of micro-post. Figure 4(e) shows the optical and scanning electron microscopy images of micro-post created using polyacrylic acid polymer. Once the droplets are deposited on the (3-Aminopropyl)triethoxysilane (APTES) coated glass slide, it is kept at normal environment for curing. After curing, polyacrylic micro-post of diameter 875 µm and height of 2 µm was obtained. This result proves the versatility of the drop impact printing technique to print micron size polymeric micro-post. Not only it reduces the processing time, but also it is cost-effective and provides more flexibility.

**Printing for Electronic Applications**

Conducting lines were printed using aqueous solutions of silver-ink and poly(3,4-ethylenedioxythiophene) polystyrene sulfonate (PEDOT:PSS) polymer. Formation of a line requires deposition of subsequent droplets at an optimum displacement. Too close a placement can lead to pattern widening, whereas too far a placement of subsequent drops will lead to discontinuity. The process is shown in Supplementary Figure S9. The line was printed using sieve #0.009 and a droplet spacing of 150 µm – 200 µm. The droplet after it touches the substrate first spreads and then oscillates. Combined effect of spreading and oscillation ensures the merging with the neighboring droplet after it lands. The concentration of silver-ink was first optimized to get good conductivity with single layer printing (Supplementary Figure S10). At the optimized silver concentration of 4% (v/v) further printing demonstrations were shown.

Figure 5(a) shows the silver line of width 450 µm, length 2.5 mm and average height of 0.655 µm. Figure 5(b) shows the PEDOT: PSS line with dimension 450 µm x 2.5 mm x 2.1 µm. Magnified image shows proper curing of polymer. The resistance of silver was found to be 31 Ω and for PEDOT: PSS was 2.7 kΩ (Figure 5(c)). This optimization plays an important role in printing-based applications and varies for different printing liquids. With this understanding, a diode was made using silver and PEDOT: PSS line printed on a glass substrate. Figure 5(d.1) and Figure 5(d.2) show the schematic diagram of the device and SEM image of the junction respectively. The IV characteristics in Figure 5(d.3) show the diode characteristics of the fabricated device.

Finally, interesting demonstrations including printed connections for LED on a flexible tape (Figure 5(e)), large area droplet array (Figure 5(f), Video SV6) and printing of letters on a flexible substrate (Figure 5(g)) are presented. Additionally, we demonstrate the possibility of scaling the printing process through multiple drop impacts on a single sieve #0.009 (Video SV7). The main advantage of drop impact printing is easy handling and cost-effective large-area printing (Figure 5(f), Supplementary Figure S11).

## Conclusion

In conclusion, this work presents a new drop on demand printing technique that as a simple design and hence requires low setup cost. Use of a superhydrophobic sieve instead of complex nozzle further reduces operational cost. Recoil ejection driven by the cavity collapse singularity leads to satellite free ejection of single droplets. The technique is found to generate monodisperse droplets. Further, this technique is able to handle a wide variety of printing solutions for different applications. As the contact between the sieve and the liquid is only for a limited duration of impact, this technique excels in printing of large particles and suspensions with high mass loading. It does not require any electric, magnetic or wave forces except a pump that will pump the liquid. This work presents an easily accessible approach to generate pico-litre to micro-litre volume droplets for different applications like bio-culture, electronic print, functional material structuring etc.

## Methods

### Nanowires Fabrication (Superhydrophobic Sieve)

Copper sieve of different pore openings and wire diameters were purchased from Copper TWPinc, USA. The growth of nanowires on copper surface was achieved by immersing the copper for 15 minutes in an aqueous solution of 2.5 *mol L$^{-1}$* sodium hydroxide and 0.1 *mol L$^{-1}$* ammonium persulphate at room temperature.[32] The nanostructured surface was further dipped in 1*H*,1*H*,2*H*,2*H*-perfluorooctyltriethoxysilane solution overnight to achieve superhydrophobicity having water contact angle ∼159° and CAH < 5° (Supplementary Fig. 12).

### Experimental Setup

The printing setup is shown in Supplementary Fig. 13. The superhydrophobic copper sieve of different pore openings (76.2μm to 533.4μm) having an area of 6 cm$^2$ was clamped from both the ends. The high-speed imaging was performed (Photron FastCam) from one side keeping the diffused led light source opposite to it. The impacting droplet was generated from an 1 mL syringe using a syringe pump, generating droplets of size 2.55 ± 0.5 mm. Teflon or APTES coated glass slides are used to collect the ejected droplet underneath the mesh at a distance of 1 mm.

## Acknowledgements

Authors would like to acknowledge Kritank Kalyan, Project assistant, CeNSE, IISc Bangalore for MATLAB Code and Solidworks assistance. A.T. would like to thank Dr. Thomas Schutzius, D-MAVT, ETH, Zurich for his valuable inputs. All the authors would like to thank Department of Science and Technology and Ministry of Electronics and Information Technology, Government of India for the financial support.

## Authors Information

Corresponding Author

* To whom correspondence should be addressed: Prosenjit Sen (P.S.), prosenjits@iisc.ac.in

[2]Present Affiliation - Australian Institute of Bioengineering and Nanotechnology, The University of Queensland, Brisbane, Australia, 4072.

[3]Present Affiliation - Department of Mechanical and Process Engineering, ETH Zürich, Sonneggstrasse 3, CH-8092 Zürich, Switzerland.

## Authors' contributions

P.S., A.K. and A.T. designed research; C.D.M., A.K. and A.T. performed research; C.D.M. and A.K. analyzed data; C.D.M., A.K., A.T. and P.S. setup model and wrote the manuscript.

## Competing interests

The authors declare no competing financial interests.

**Figure 1: Drop impact technique: Mechanism and explanation.**

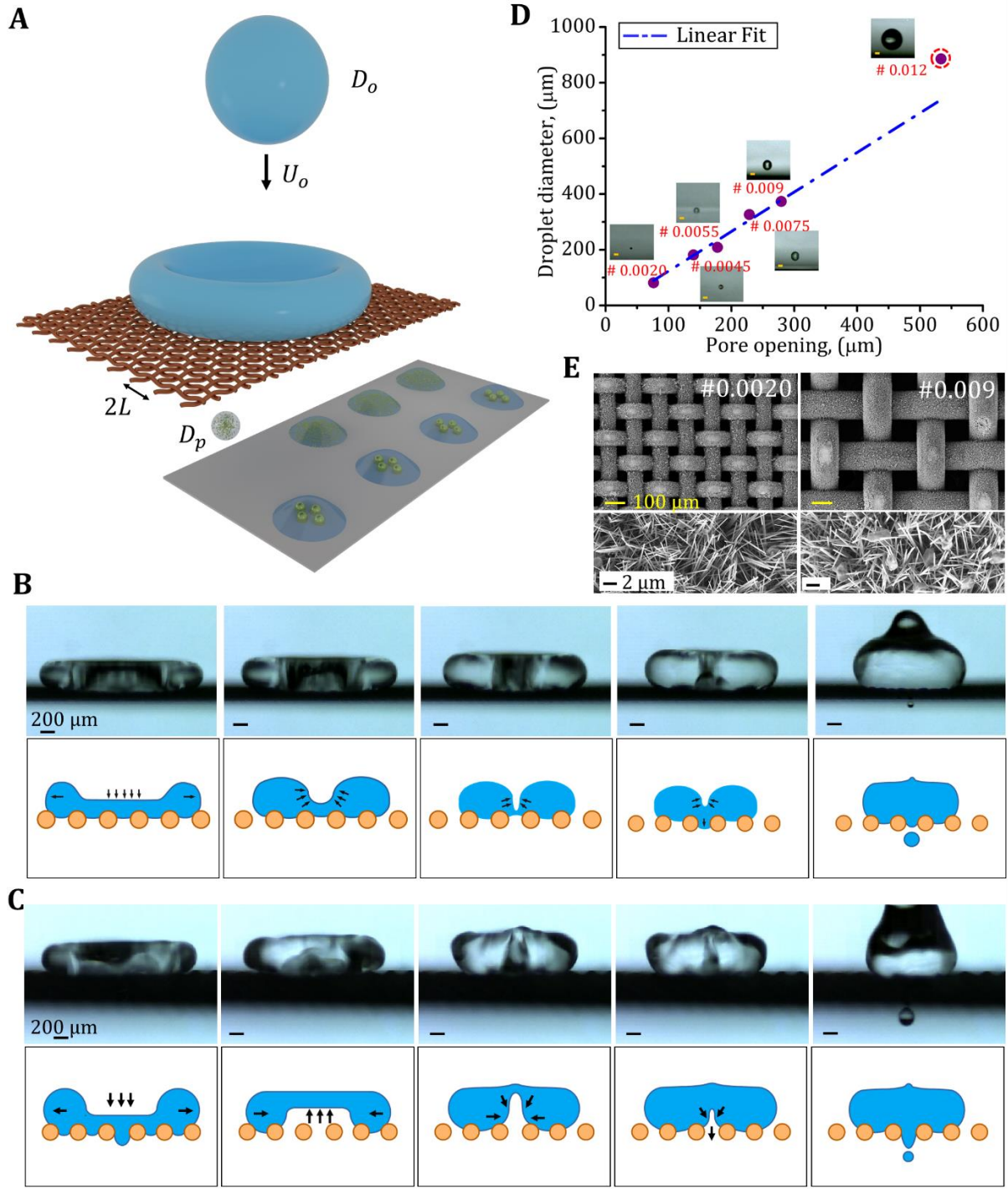

(a) Schematic illustration showing drop impact setup, droplet (diameter $D_o$, Velocity $U_o$) impacting on a superhydrophobic sieve (pore opening, $L$) to eject out a single smaller droplet (diameter $D_p$). The impacting drop gives rise to two modes of single droplet ejection, (b) Impact cavity and (c)

Recoil cavity. The time-lapsed images and schematic illustration for impact cavity and recoil cavity modes show the mechanism of cavity formation and collapse using sieve #0.0045 with 40% glycerol water droplet and sieve #0.009 with pure water droplet respectively. The drop impact technique explored in terms of smallest ejected droplet that can be generated, (d) showing as a plot between water droplet diameter versus pore opening and insets showing corresponding patterned droplet (Scale bar - 100μm). Superhydrophobic sieves with different pore openings were used starting from sieve type #0.012 to #0.0020 (L-76.2μm, W-50.8μm). (e) SEM images of sieves #0.009 and #0.0020.

**Figure 2: Parametric studies showing the capabilities of drop impact (DI) printing technique.**

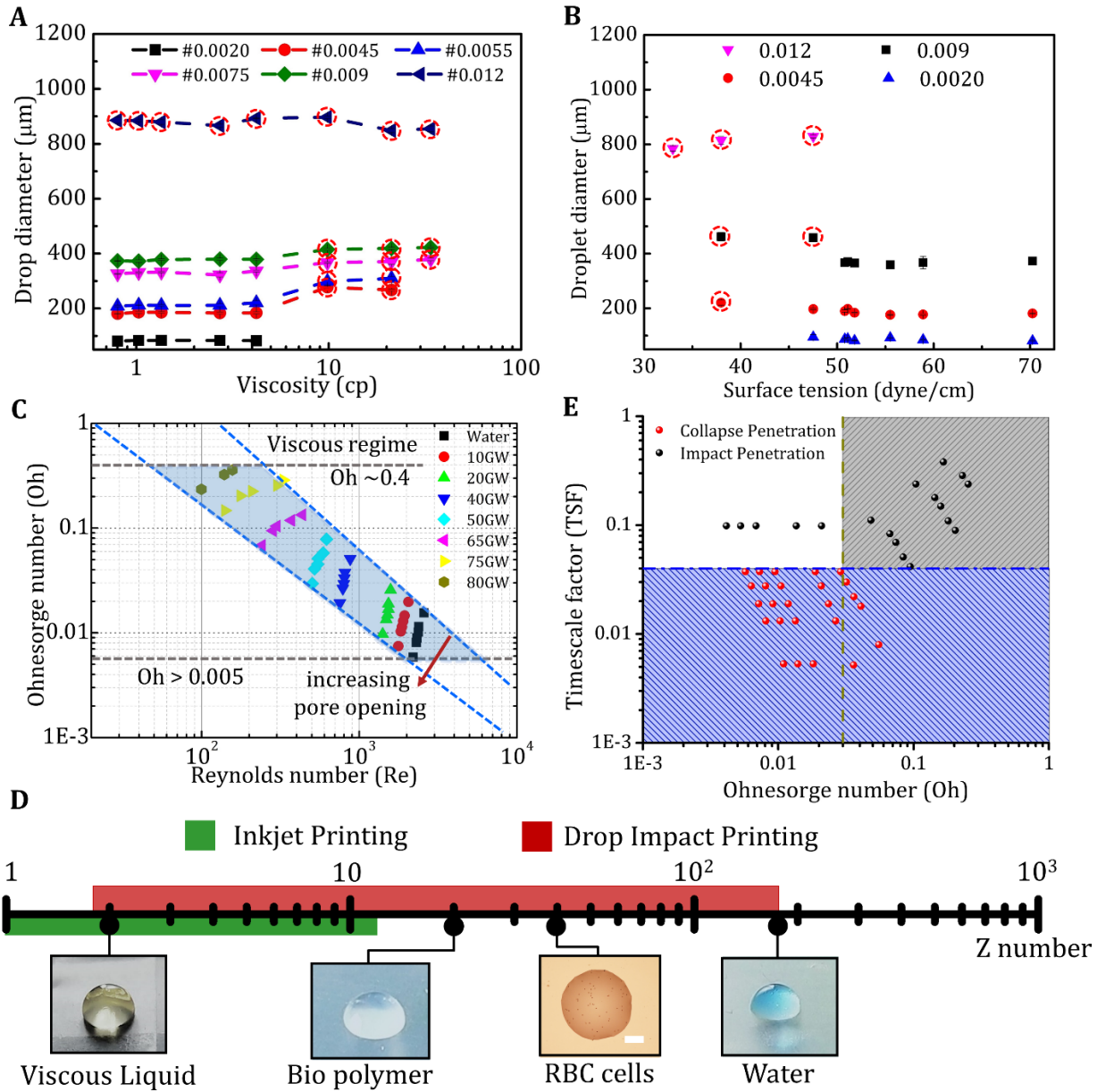

The extent of viscous liquid and low surface tension liquid printing was explored using glycerol water solution, PEG water solution and ethanol-water solution. The ejected droplet diameter was plotted with (**a**) liquid viscosity and (**b**) liquid surface tension for sieve with different pore openings. (**c**) The printable regime was observed in the plot between Ohnesorge number and Reynolds number. The light blue shaded part shows the printable region of drop impact printing technique. The range gives us an idea of the extent of different liquids that can be used for printing. (**d**) The broad range of liquids are shown in the terms of Z number with inset images showing the different liquids drop that can be printed. The drop impact technique was compared to inkjet printing with red and green bar for drop impact and inkjet printing respectively. (**e**) The mechanism of different ejection mode was explained based on timescale factor with varying Ohnesorge number. The critical Ohnesorge number that ensures transition from inertial to viscous regime was 0.03 and time scale factor value that defines the transition from CPM to IPM was found to be 0.04.

**Figure 3: Clogging free printing: Large particle size and higher mass loading printing.**

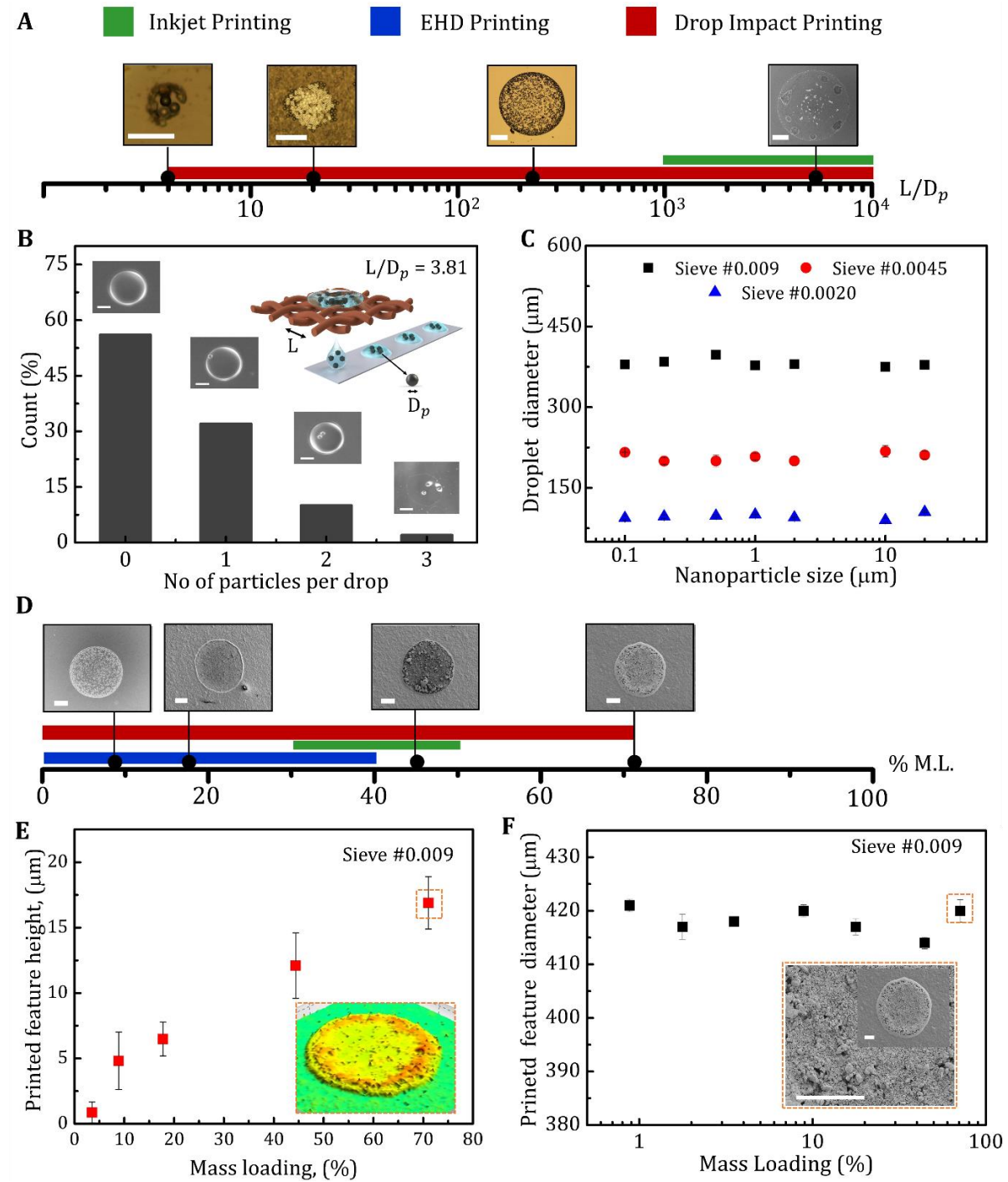

The clogging free printing was demonstrated based on ability to print large particle size and high mass loading suspensions. (**a**) The larger particle size printing ability was shown in linear $L/D_p$

Chart with inset showing different printed particle size for different $L/D_p$ (Scale bar -100μm) ratios. $L/D_p$ can be as low as 3.81 for drop impact printing which is significantly smaller as compared to inkjet printing. (**b**) The percent count to print a single bead and multiple beads in a drop is demonstrated. The probability of single bead capturing in a single drop (80 *μm* diameter) was found to be 32%. The inset shows number of beads in a single drop (Scale bar – 100 *μm*). (**c**) Further the printed droplet diameter with varying particle size was shown. The droplet diameter was independent of different particle size suspension. (**d**) The linear chart shows as high as 71% mass loading suspension solution printing is possible using drop impact printing as compared to inkjet and electrohydrodynamic (EHD) printing. Inset showing SEM image of printed droplet for different mass loading (Scale bar -100 *μm*). (**e**) The printed feature height was shown with varying mass loading. The increase in drop height with increase in mass loading gives us a way to print particle suspensions in one go. (**f**) Further printed feature diameter was plotted with varying mass loading (Scale – 100 *μm*). The printed drop size was found to be independent with increase in mass loading. Insets in both figure (e) and (f) show higher mass loading printed drop.

**Figure 4: Drop impact Printing of Biological solutions and Biopolymer.**

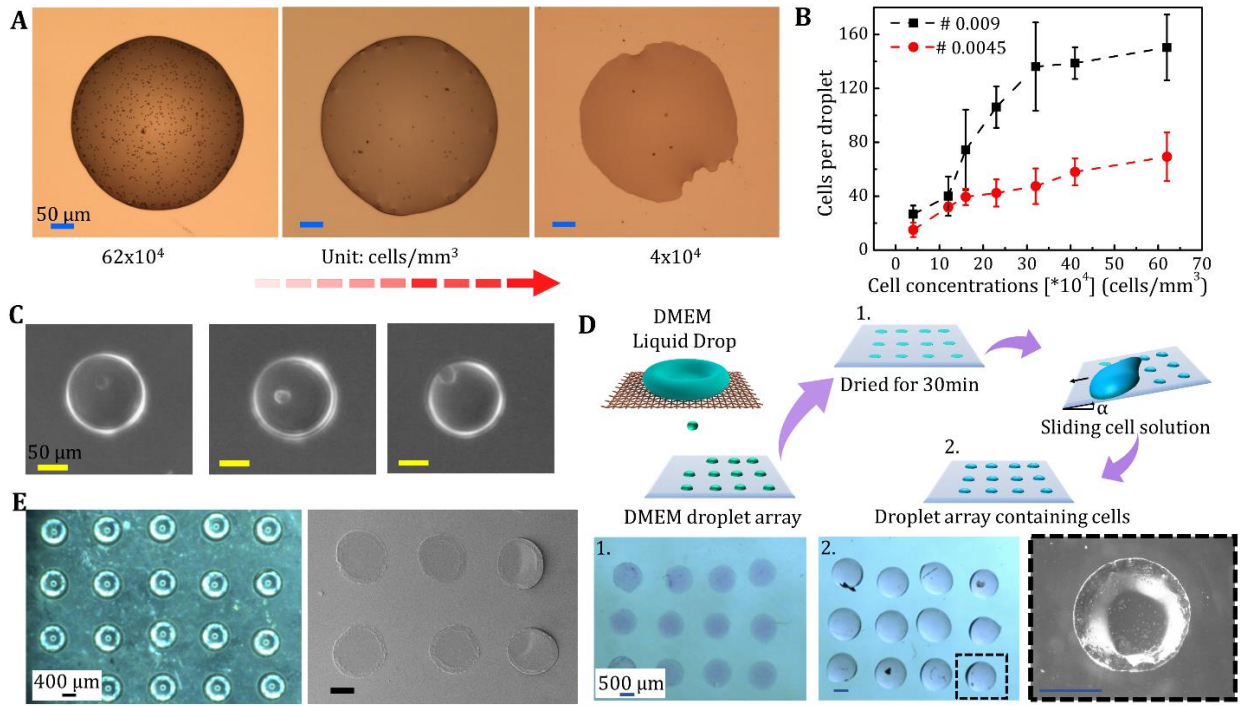

(**a**) Microscopic images of single droplet patterned using cell (RBC) laden PBS solution of different concentrations, (Scale bar – 50 *μm*). The cells are contained in an isolated droplet of volume 26 *nL* patterned using mesh type #0.009. (**b**) The number of cells per droplet for varying cell concentrations was examined for mesh-type #0.009 and #0.0045. The single cell printing was further demonstrated using drop impact technique. (**c**) Single cells (MDAMB 231) of average size ~17 *μm* trapped in 0.268 *nL* single drop. The drops were collected on an oil coated glass slide. The concentration of cells solution was kept at $50*10^4$ *cells/mm³* (Scale bar - 50 *μm*). (**d**) Illustration showing printed DMEM droplet arrays using drop impact technique. (1) shows printed DMEM droplets on hydrophobic Teflon surface. (2) shows the arrays of cells containing droplets after cells solution swipe and magnified image of a printed droplet containing cells. Beside this, the technique ability was explored by using bio polymeric viscoelastic liquid (0.0125*gm/ml* polyacrylic acid mixed in water) for 3D printing applications. (**e**) The large patterned micro-posts of 875 *μm* diameter and 2 *μm* height was printed on APTES coated glass slides and the corresponding SEM image (Scale bar - 200 *μm*).

**Figure 5: Printing of electrically conducting materials for large-area fabrication and flexible electronics applications.**

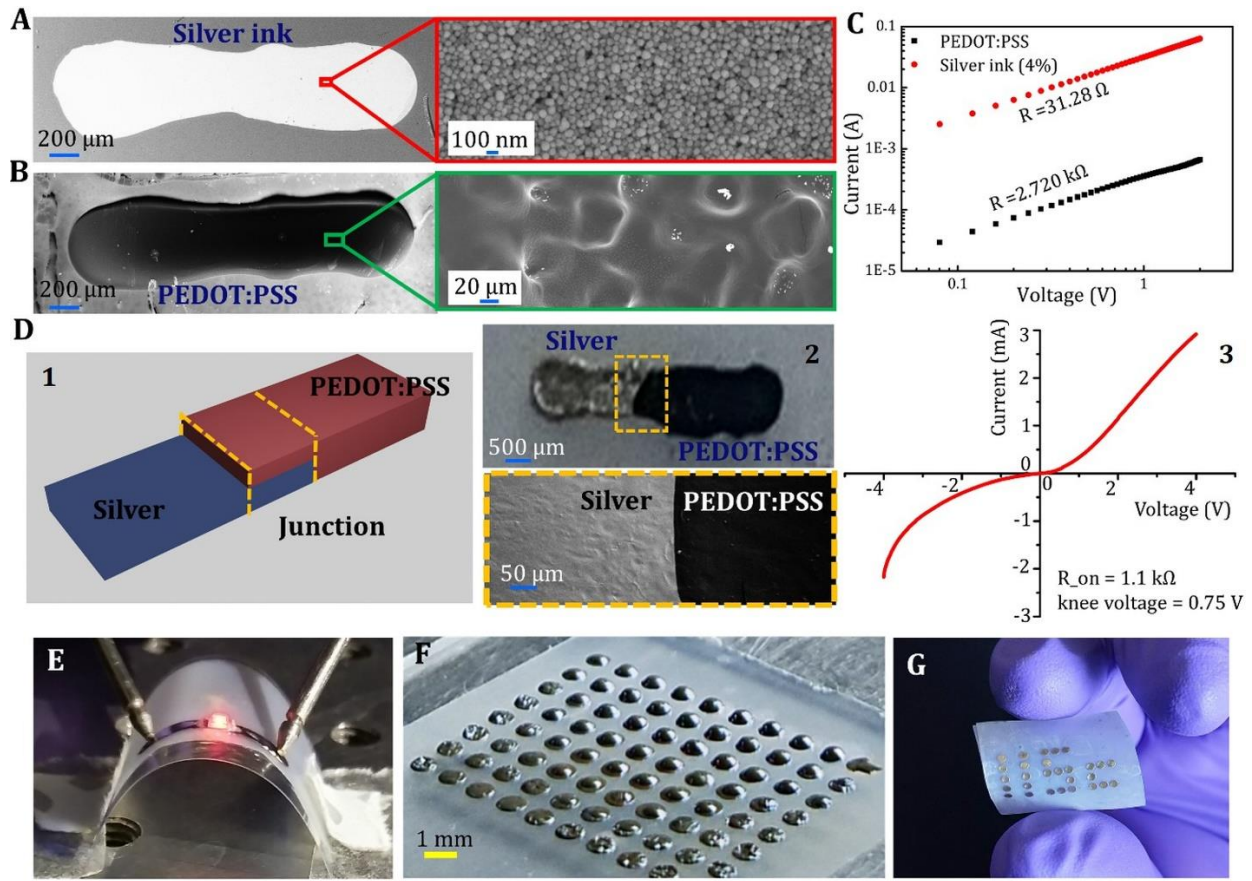

Room temperature printing of (**a**) silver ink (4% (v/v)) conductive line and the corresponding SEM image. (**b**) PEDOT: PSS printed line and the SEM image showing the connectivity. (**c**) IV characteristics of both silver and PEDOT: PSS conducting lines. (**d**) (1) Silver ink and PEDOT: PSS further used to form a junction to show the capability of the technique for electronic applications. (2) Optical microscopic image and the SEM image showing the junction. (3) Additionally, IV characteristic was performed for the junction to check the connectivity. Further, as a demonstration (**e**) two silver conducting lines are connected using drop impact printing technique and the voltage is applied at both ends to show the glowing LED. (**f**) Large area droplet patterning and (**g**) flexible printing were also shown to demonstrate the wide applicability of the technique.

**Supplementary Information for**

# Drop Impact Printing

Chandantaru Dey Modak[1], Arvind Kumar[1,2], Abinash Tripathy[1,3] and Prosenjit Sen[1,*]

[1]Centre for Nano Science and Engineering, Indian Institute of Science, Bangalore, India, 560012

**Corresponding Author's Email:** prosenjits@iisc.ac.in

[2]Present Affiliation - Australian Institute of Bioengineering and Nanotechnology, The University of Queensland, Brisbane, Australia, 4072.

[3]Present Affiliation - Department of Mechanical and Process Engineering, ETH Zurich, Sonneggstrasse 3, CH-8092 Zürich, Switzerland.

**This PDF file includes:**
Supplementary text
Figures S1 to S13
Tables S1 to S4
Captions for Video S1 to S7

**Other supplementary materials for this manuscript include the following:**

Video S1 to S7.

**Supplementary Information Text**

**Printing Solution Preparation**

**Glycerol-water Solution and ethanol-water solution**

Glycerol and ethanol were purchased from SD Fine Chemicals, Bangalore, India. Different concentrations (v/v %) of glycerol-water and ethanol-water solutions were prepared for the experiment. The glycerol water concentration was varied from 10% to 80% and for the ethanol-water solution, it was from 12% to 36%. The solution viscosity, surface tension and density are measured using Rheolab QC rheometer from Anton Paar, Density meter: DMA™ 4200 M from Anton Paar, and Tensiometer: K20 from Kruss Scientific respectively. The liquid properties of glycerol-water solutions and ethanol-water solutions are listed in Table S3A & 3B.

**PEG-water solution**

Polyethylene glycol 4000 was purchased from Sigma Aldrich. PEG of different weight% was mixed in 50 mL of distilled water and stirred, till it completely dispersed. The concentration was varied from 1% to 10% (v/v). The liquid properties like surface tension, density and viscosities are measured. The liquid properties of PEG-water solutions are listed in Table S3C.

**Nanoparticle Suspensions**

Different nanoparticles of varying sizes (10 nm to 20 µm) were used in the experiment and are purchased from US Nanomaterials Research, Inc. Nanoparticles of different weight were mixed in 50 mL of 10% (v/v) PEG-water solution and stirred for 30 minutes. In order to have dispersed solutions, suspensions are further sonicated for 1 hour before the experiments. For nanoparticle size variation demonstration, the concentration of nanoparticles was fixed around mass loading of 8.88% (w/w). And for different mass loading demonstrations, 200 nm Zirconium dioxide nanoparticles were used and the concentration (mass loading) was varied from 0.88% to 71%. The different nanoparticles size specifications are listed in Table S4.

**Electronic Inks**

For electronic ink printing applications two inks are used in the present study: poly(3,4-ethylenedioxythiophene) polystyrene sulfonate (PEDOT: PSS) and silver ink. PEDOT: PSS (1.3 wt % dispersion in $H_2O$) and silver ink (30-35 wt%) were purchased from Sigma Aldrich. The aqueous silver ink solutions of varying concentration (1% to 4%, v/v) were prepared by mixing it in 10% (v/v) PEG-water solution. The ink suspensions were sonicated for 1 hour before the experiments.

**Polymeric Solutions**

Polyacrylic acid (PAA) was purchased from Sigma Aldrich. Polyacrylic acid of 0.5 gm was mixed in 40 mL of distilled water and then stirred for 1 hour. The prepared viscous mixture was used for printing applications. The viscosity of the polymeric solution was measured to be 1.15 mPas.

**Table S1:** Sieve properties

| Mesh Type [#] | Pore opening, mm | Wire diameter, mm | % Opening |
|---|---|---|---|
| 0.012 | 0.5334 | 0.3048 | 40 |
| 0.009 | 0.2794 | 0.2286 | 30 |
| 0.0075 | 0.2286 | 0.1905 | 30 |
| 0.0055 | 0.1778 | 0.1397 | 31 |
| 0.0045 | 0.1397 | 0.1143 | 30 |
| 0.0020 | 0.0762 | 0.0508 | 35 |

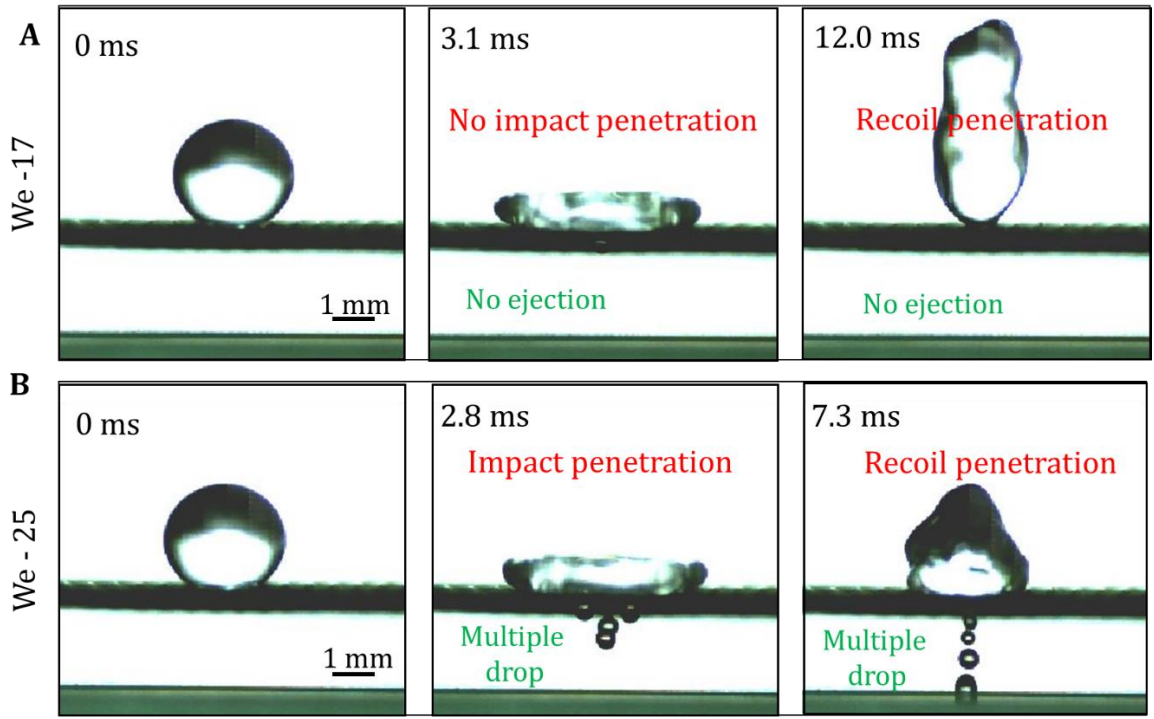

**Supplementary Figure S1:** Time-lapse images of water droplet impacting on superhydrophobic sieve (#0.009) from different heights. (**a**) The drop impacts from a height of 2.5 cm (We-17), resulting in neither impact penetration nor recoil penetration (**b**)

The drop impact from height of 3.7 cm (We-25), resulting in multiple droplets ejection from single jet. Droplet ejection was observed both in impact and recoil penetration.

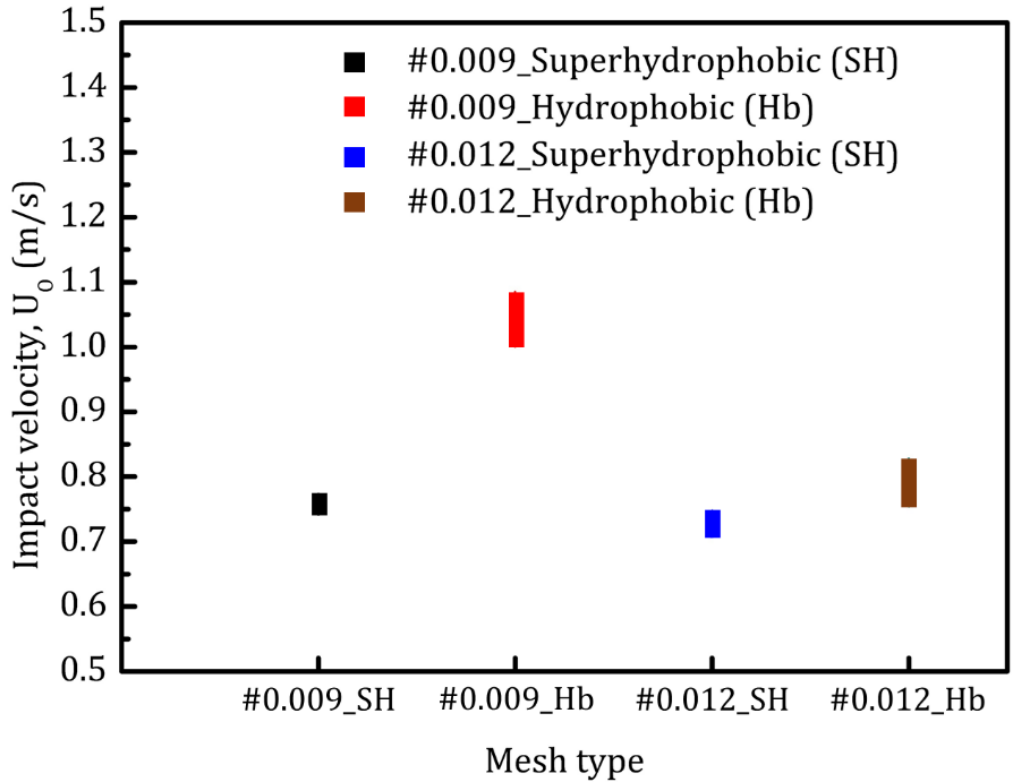

**Supplementary Figure S2:** The comparison of single drop regime for superhydrophobic and hydrophobic sieve (#0.009 and #0.012) The plot shows the range of impact velocities that ensures single drop ejection when water is used as printing liquid.

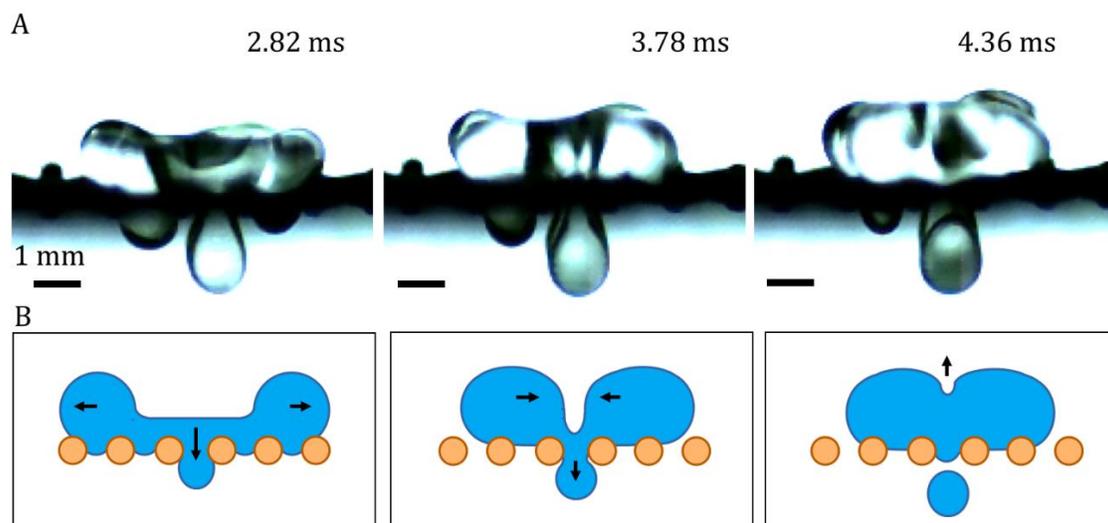

**Supplementary Figure S3:** The single drop ejection mechanism for sieve type #0.012 is shown using (**a**) time lapse images and (**b**) schematic illustration. When droplet impacts on sieve, impact jet is formed. As the drop starts to recoil, impact jet is not able to retract back completely and recoil jet pushes the jet further to eject single drop. In this case, the collapse of top interface of the drop is responsible for single drop ejection.

**Simulation Results**

Drop impact simulation performed using the phase field approach in COMSOL. Flow field and phase field were solved in cylindrical coordinate system to reduce the computational time. As it is impossible to model the sieve in 2D cylindrical coordinate system, we simplified our simulation by modelling only for the central pore. The pore dimension and the wire size are matched with that of sieve #0.009. We modelled impact of water droplet with radius of 1250 µm with an impact velocity of 0.8 m/s. Though the simulation was able to capture several aspects of the impact phenomenon, but we did not observe formation of recoil cavity. Though the liquid in the pores was observed to recoil back, the amount of fluid recoiling back was insufficient for formation of recoil cavity. This was because only one pore was simulated. The color scheme shows pressure with red representing higher values and green representing lower values. The arrows show flow direction. Size of the arrows represent the magnitude of local velocities. The images on the right are snapshots from a high-speed video for an impact on sieve #0.009.

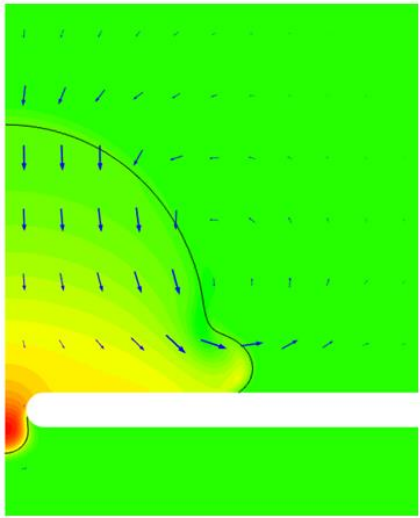
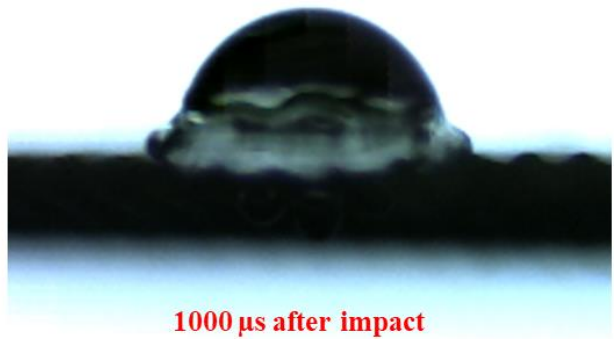

1000 µs after impact   1000 µs after impact

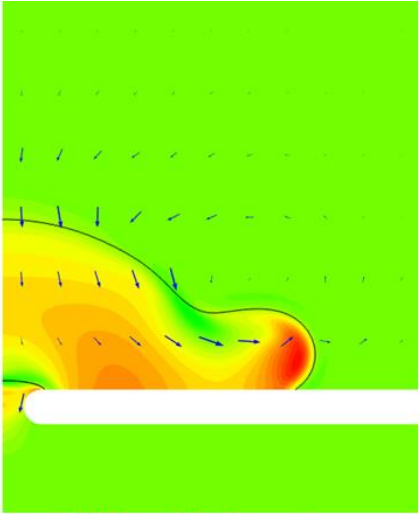
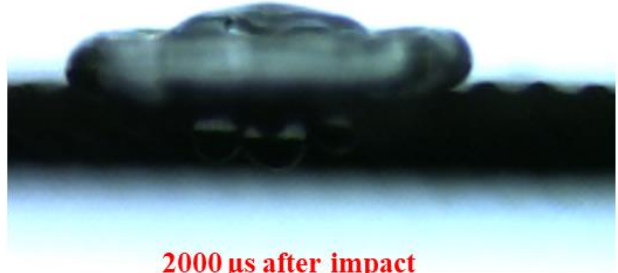

2000 μs after impact

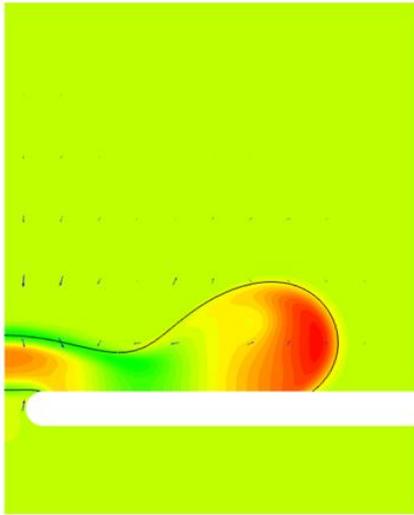
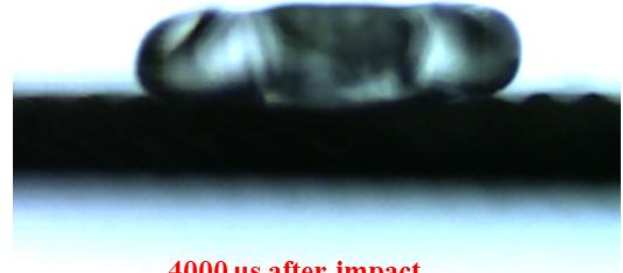

4000 μs after impact

As the simulations do not capture the recoil cavity formation (due to structural difference with the experimental condition), the dynamics start deviating in later stages.

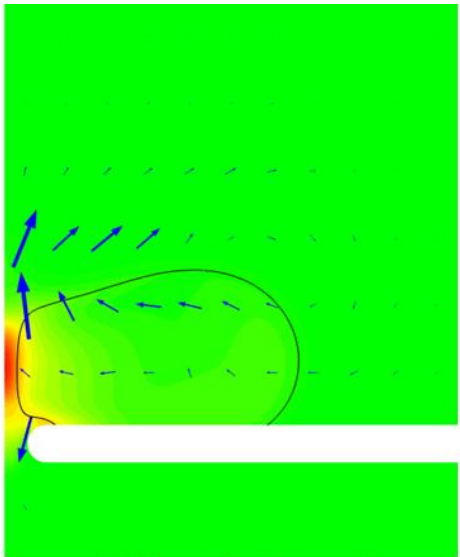
**6000 μs after impact**

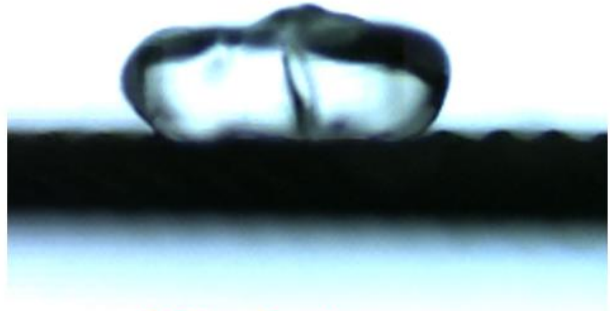
**5134 μs after impact**

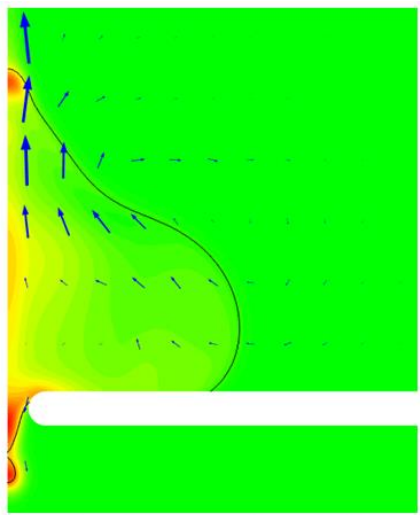
**Microdroplet separation at 6860 μs**

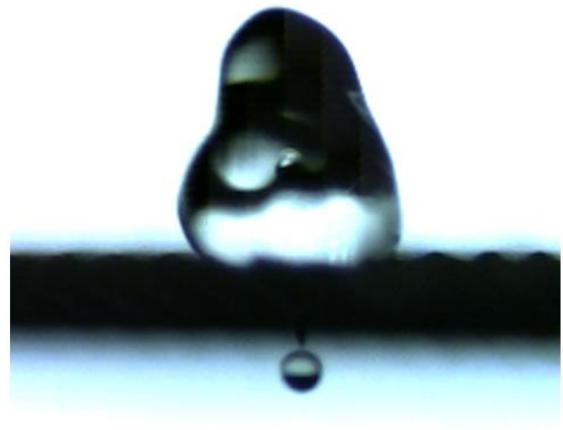
**6634 μs from impact**

Despite differences in the simulation and experiments, microdroplets are generated approximately at the same time.

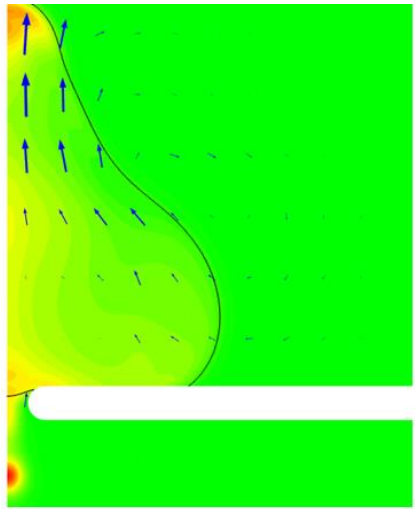 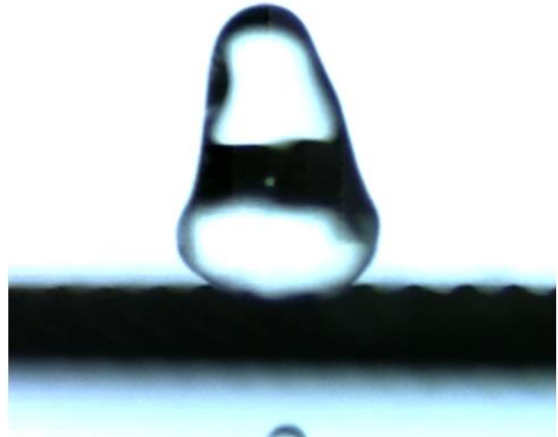

7400 μs after impact                    7400 μs after impact

**Supplementary Figure S4:** Simulation and experimental result at different time scales showing the single drop ejection.

**Supplementary Table S2:** Dimensionless numbers of water droplet printing

| Mesh Type | Impact height, cm | Impact Velocity, cm/s | Weber number, Re | Reynolds number, Re | Ohnesorge Number, Oh |
|---|---|---|---|---|---|
| 0.012 | 2.55 | 70.718 | 17.74 | 2201.4 | 0.0058585 |
| 0.009 | 2.8 | 74.104 | 19.48 | 2306.8 | 0.0080947 |
| 0.0075 | 2.85 | 74.762 | 19.83 | 2327.3 | 0.008949 |
| 0.0055 | 2.95 | 76.063 | 20.52 | 2367.8 | 0.0101472 |
| 0.0045 | 3.00 | 76.705 | 20.87 | 2387.7 | 0.0114476 |
| 0.0020 | 3.45 | 82.257 | 24.00 | 2560.5 | 0.0155002 |

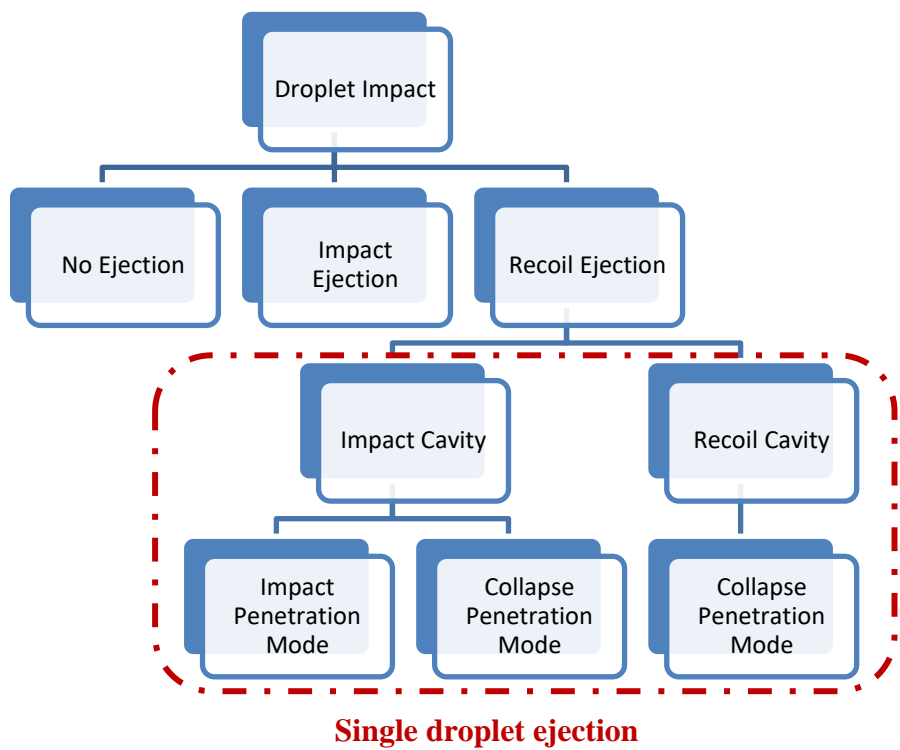

**Supplementary Figure S5:** Different collapse dynamics resulting from impact of a drop on sieve to generate single drop is shown. The single drop ejection was possible only in recoil ejection. Under recoil ejection, further distinction was made based on cavity collapse modes.

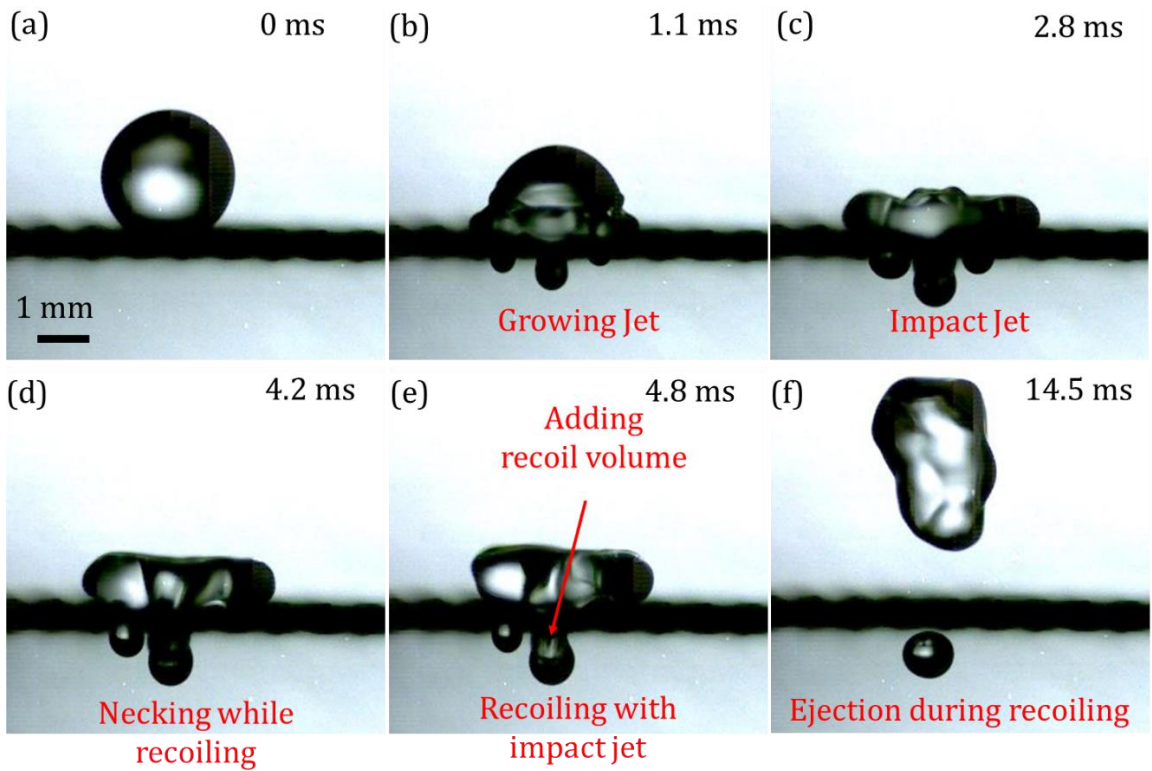

**Supplementary Figure S6:** Time lapse images of water drop impacting on sieve (#0.012). This ejection mode comes under impact penetration mode. In this mode, impact jet contributes to single droplet volume since it is not able to retract back to parent drop completely.

**Table S3:** Fluid properties

**A. Newtonian Fluid** – Aqueous Water Glycerol Solution

| Liquid Code | Concentration (v/v)% | Density, kg/m$^3$ | Viscosity, mPas | Surface Tension, N/m |
|---|---|---|---|---|
| W | 0 | 997.08 | 0.8007 | 0.07025 |
| 10GW | 10 | 1020.7 | 1.03 | 0.06993 |
| 20GW | 20 | 1045.25 | 1.35 | 0.06949 |
| 40GW | 40 | 1097.1 | 2.72 | 0.06835 |
| 50GW | 50 | 1123.75 | 4.21 | 0.06762 |
| 65GW | 65 | 1164.75 | 9.85 | 0.06668 |
| 75GW | 75 | 1191.95 | 21.2 | 0.06535 |
| 80GW | 80 | 1205.45 | 33.9 | 0.06482 |
| 85GW | 85 | 1218.7 | 58 | 0.06426 |

**B. Newtonian Fluid** –Ethanol Water Solution

| Liquid Code | Concentration (v/v)% | Density, kg/m$^3$ | Viscosity, mPas | Surface Tension, N/m |
|---|---|---|---|---|
| 12EW | 12.4 | 888.07 | 0.8405 | 0.04753 |
| 24EW | 24.5 | 864.40 | 0.671 | 0.03797 |
| 36EW | 36.2 | 839.97 | 0.523 | 0.03298 |

**C. Non-Newtonian Fluid** – Aqueous PEG Solution

| Liquid Code | Concentration (v/v)% | Density, kg/m$^3$ | Viscosity, mPas | Surface Tension, N/m |
|---|---|---|---|---|
| 1PEG | 1 | 999.0791 | 0.67 | 0.0589 |
| 2PEG | 2 | 1000.723 | 1.2 | 0.0555 |
| 5PEG | 5 | 1005.681 | 2.9 | 0.0518 |
| 7PEG | 7 | 1009.01 | 4.12 | 0.0511 |
| 10PEG | 10 | 1014.037 | 6.1 | 0.0508 |

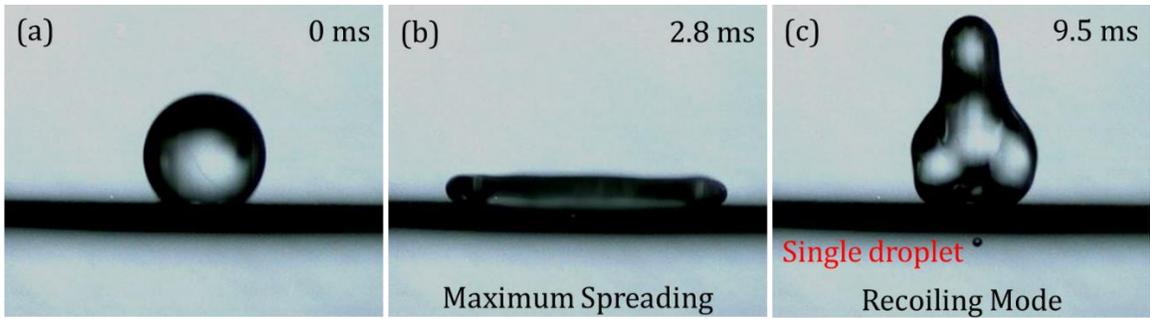

**Supplementary Figure S6:** Time-lapse images of viscoelastic drop when impacted on superhydrophobic sieve (#0.0045). The liquid used was 10% (v/v) Xanthum gum-water solution.

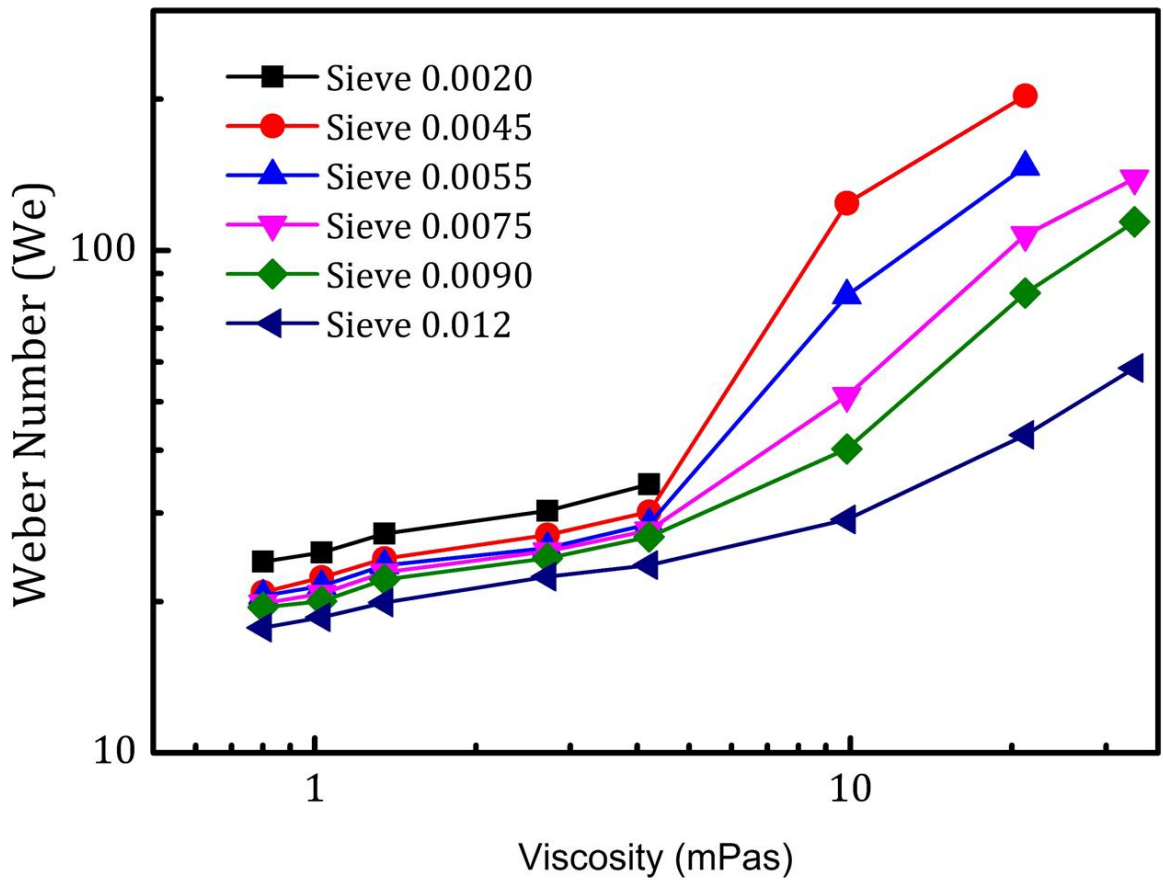

**Supplementary Figure S7:** Weber number corresponding to single droplet ejection with varying viscosity for different sieves are shown. The liquid used is glycerol water mixture of different concentrations.

**Table S4:** Different nanoparticles and size specifications.

| Particle Type | Average particle size, μm |
|---|---|
| Titanium Oxide, $TiO_2$ | 0.1 |
| Zirconium Oxide, $ZrO_2$ | .2 |
| Zirconium Oxide, $ZrO_2$ | .5 |
| Halloysite nanoclay | 1 |
| Polystyrene Beads | 10 |
| Polystyrene Beads | 20 |

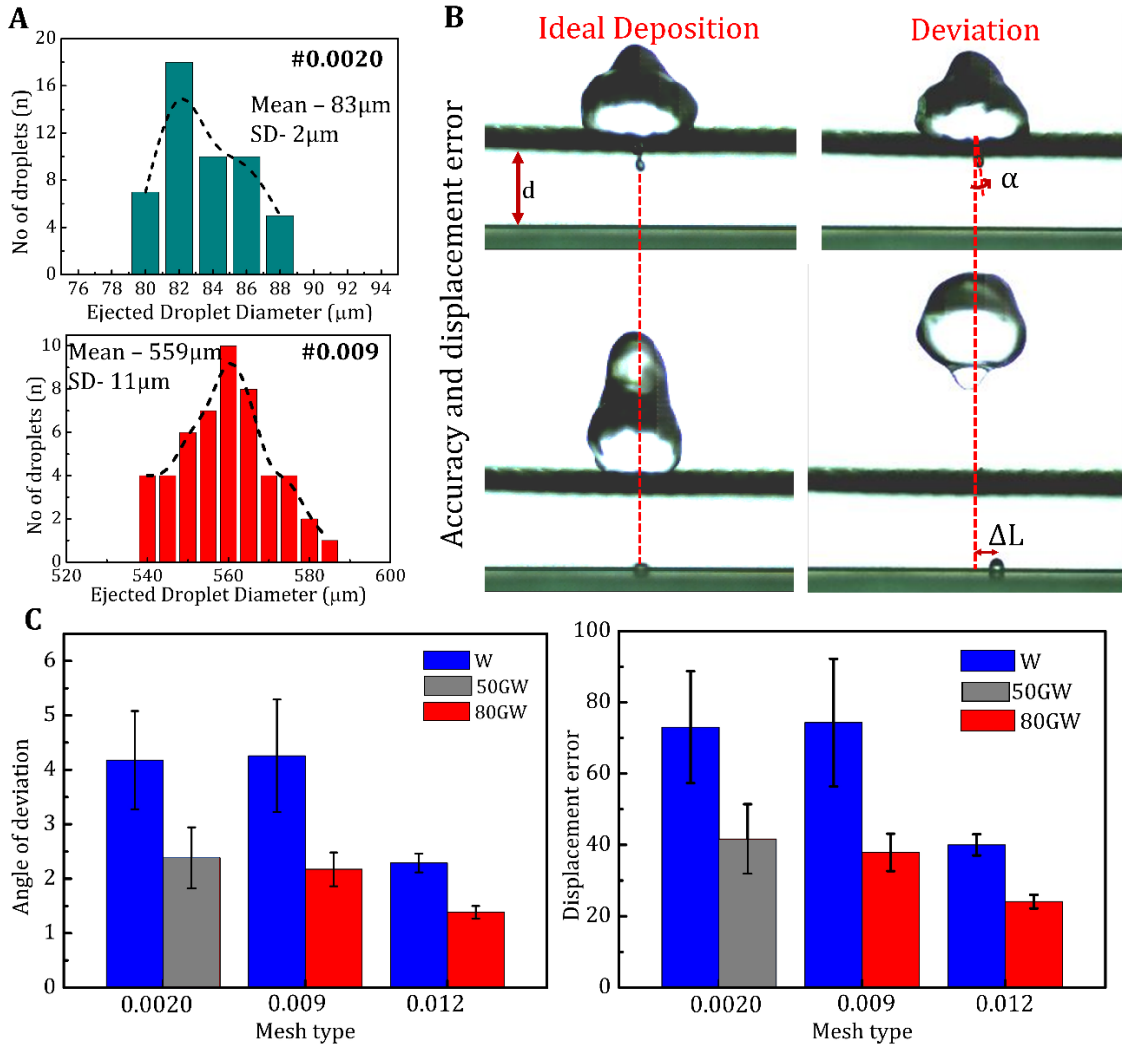

**Supplementary Figure S8:** (**a**) Printed droplet size distribution is shown for sieve #0.0020 and #0.009. (**b**) Accuracy of the technique is measured by estimating the angle of deviation and displacement error as shown. (**c**) Angle of deviation and displacement error for three different sieves and three different viscosity liquids (Water, 50% glycerol water mixture, and 80% glycerol water mixture) was measured.

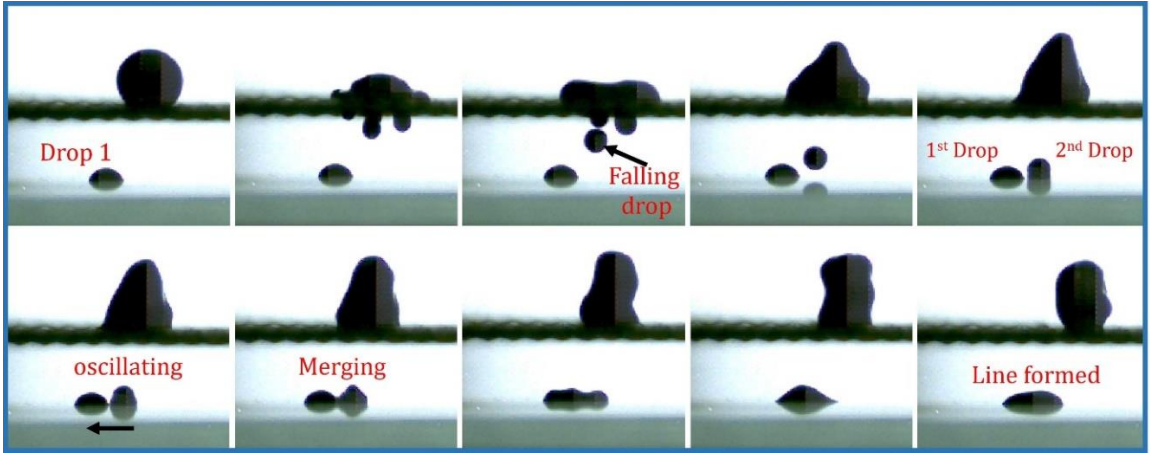

**Supplementary Figure S9:** The sequence of images showing the ejecting silver ink droplet and merging with the neighborhood drop to form a line. The printed drop oscillated and merged with the neighborhood droplet. The droplets were printed on glass slide embedded with scotch tape. The spacing between the drops was kept between 150 µm to 200 µm for printing droplet volume of 0.3 µL.

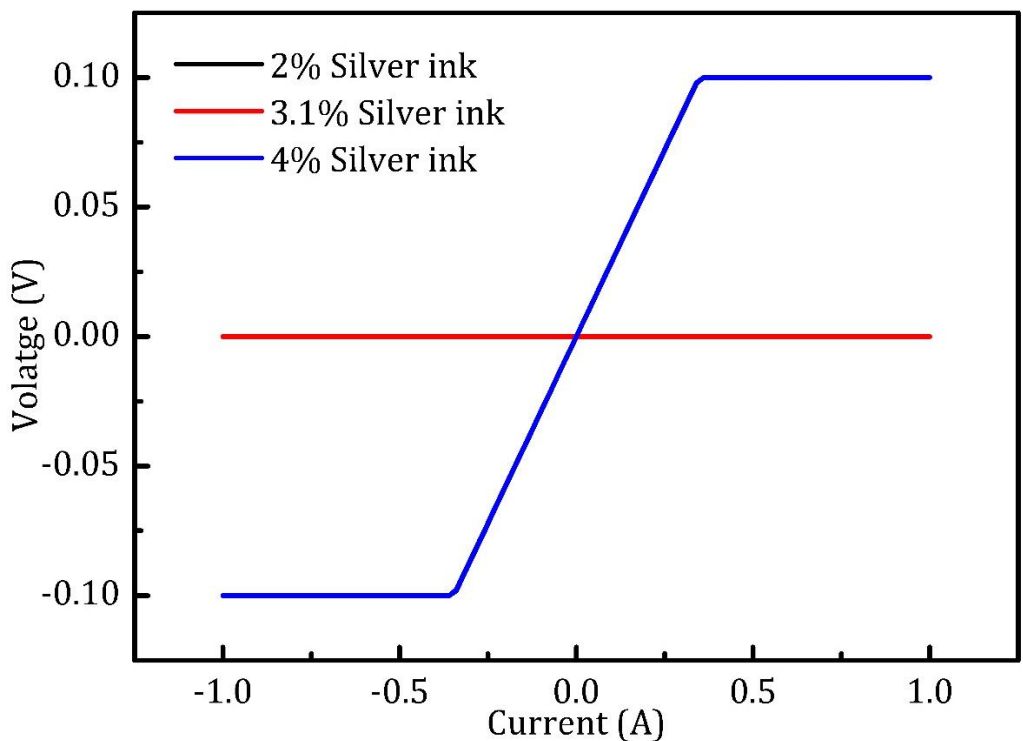

**Supplementary Figure S10:** Voltage versus current curves for different silver ink concentrations. The optimization of silver ink line was carried out using mesh type #0.009.

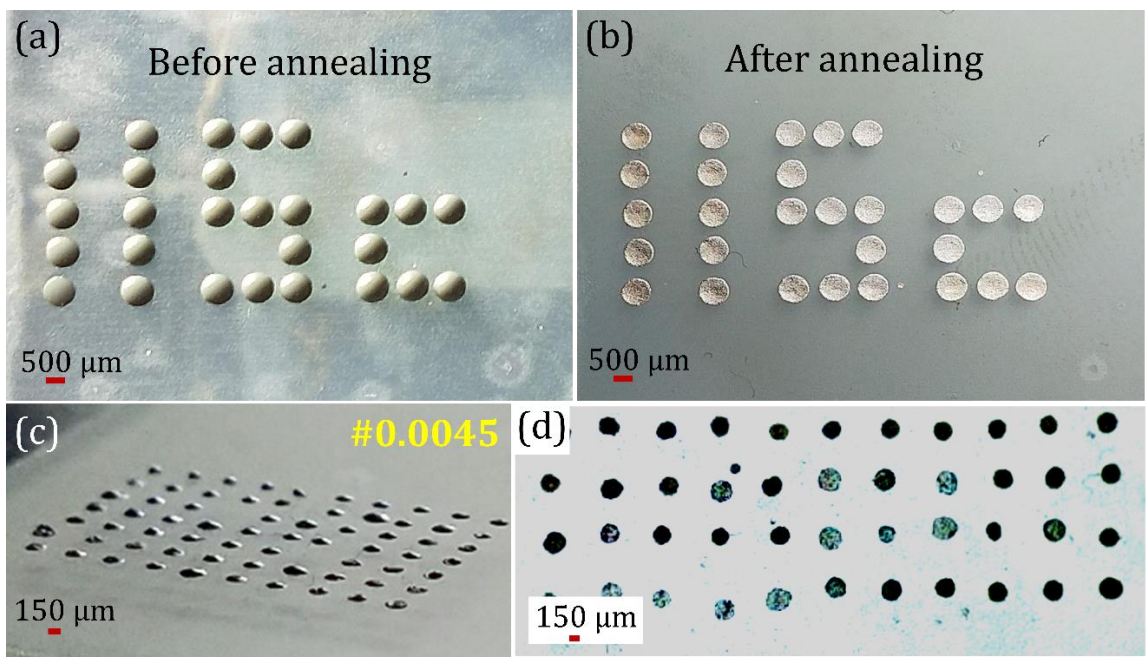

**Supplementary Figure S11:** Optical microscopy images of silver ink based printed letters using drop impact printing technique **(a)** before annealing and **(b)** after annealing. In this case printing was done with mesh type #0.012 with ejected droplet volume approximately 0.35 µL. Large area droplet array printing capability of the technique was explored using mesh type #0.0045 showing patterned silver ink droplet from different view, **(c)** side angled view and **(d)** top view. The printed droplet volume was approximately 3 nL.

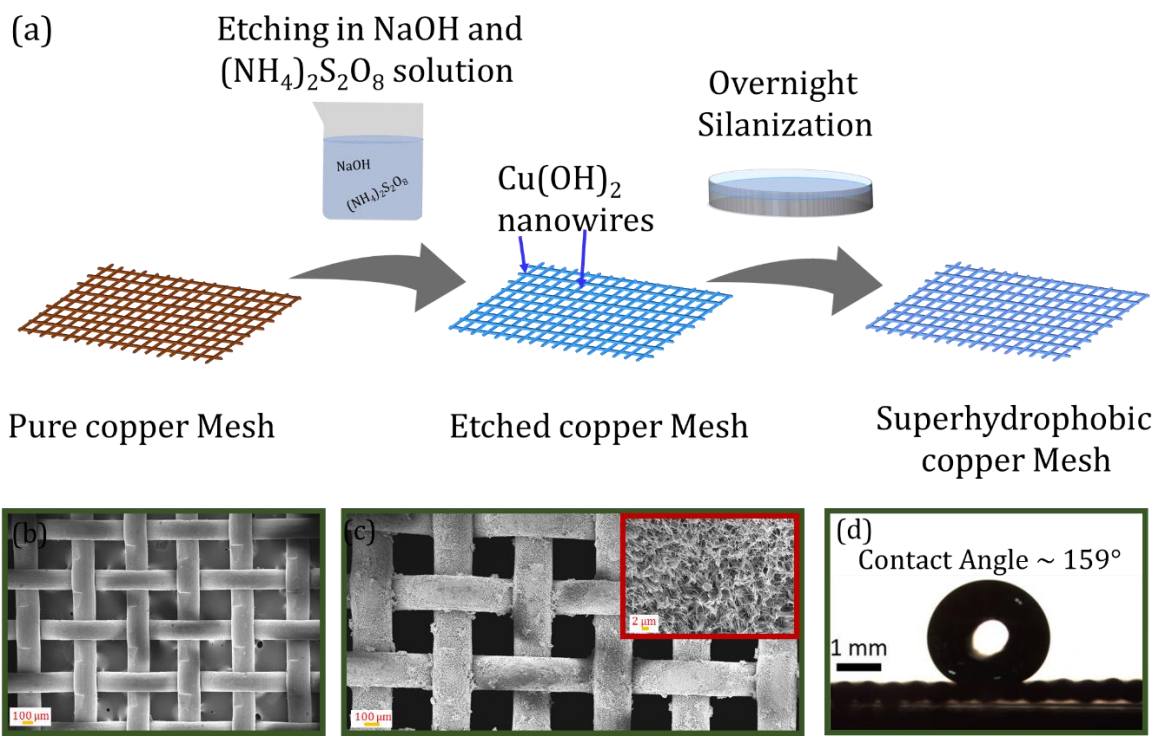

**Supplementary Figure S12:** **(a)** Schematic showing etching of cleaned copper sieve followed by silanization to obtain superhydrophobic sieve. SEM of sieve for **(b)** pure copper sieve, **(c)** etched silanized superhydrophobic sieve and inset showing the nanowires that are present on the surface. **(d)** contact angle and contact angle hysteresis measurements revealed the contact angle to be 159° and the hysteresis to be <5°.

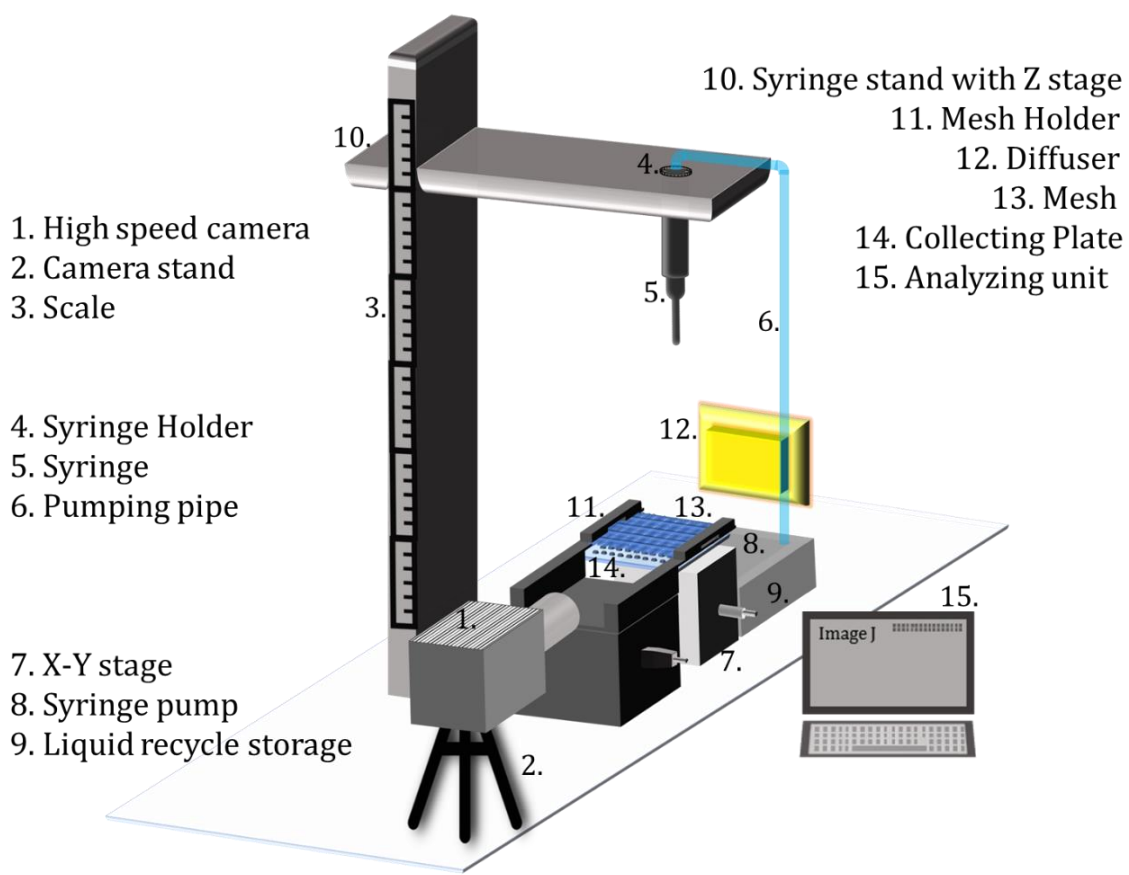

**Supplementary Figure S13:** Drop impact printing setup.

**Movie S1.** Single droplet printing during recoil ejection using sieve #0.0045.

**Movie S2.** Single droplet printing through impact cavity collapsed penetration mode.

**Movie S3.** Single droplet printing through recoil cavity collapsed penetration mode.

**Movie S4.** Single droplet printing through impact cavity impact penetration mode.

**Movie S5.** Ejected droplet diameter for different pore openings.

**Movie S6.** Impacting droplet and moving substrate underneath mesh for single droplet printing.

**Video S7.** Multiple droplets impacting and ejecting successive single droplets in a row.